\documentclass[reprint,aps,showpacs]{revtex4-1}
\usepackage{graphicx,subfig}
\usepackage{mathrsfs}
\usepackage{graphicx}
\usepackage{slashed}
\usepackage{ragged2e}
\usepackage[font=small]{caption}
\usepackage{caption}
\usepackage{color}
\usepackage{hyperref}
\usepackage{amsmath, amsfonts, amssymb, bm}
\usepackage{braket,cancel}
\begin{document}
	%%%%%%%%%%%%%%%%%%%%%%%%%%%%%%%%%%%%%%%%%%%%%%%%%%%%%%%%%%%%%%%%%
	\title{Relativistic strong-field ionization of hydrogen-like atomic systems in constant crossed electromagnetic fields}
	\author{A. Eckey$^{1}$}
	\author{M. Klaiber$^{2}$}
	\author{A. B. Voitkiv$^{1}$}
	\author{C. M\"uller$^{1}$}
	\affiliation{$^{1}$Institut f\"ur Theoretische Physik I, Heinrich Heine Universit\"at D\"usseldorf, Universit\"atsstra{\ss}e 1, 40225 D\"usseldorf, Germany} \affiliation{$^{2}$Lochhofstra{\ss}e 8, 78120 Furtwangen, Germany}
	\date{\today}
	%%%%%%%%%%%%%%%%%%%%%%%%%%%%%%%%%%%%%%%%%%%%%%%%%%%%%%%%%%%%%%%%%
\begin{abstract}
	Relativistic strong-field ionization of hydrogen-like atoms or ions in a constant crossed electromagnetic field is studied. The transition amplitude is formulated within the strong-field approximation in G\"oppert-Mayer gauge, with initial and final electron states being described by the corresponding Dirac-Coulomb and Dirac-Volkov wave functions, respectively. Coulomb corrections to the electron motion during tunneling are taken into account by adjusting an established method to the present situation. Total and energy-differential ionization rates are calculated and compared with predictions from other theories in a wide range of atomic numbers and applied field strengths.
\end{abstract}
%%%%%%%%%%%%%%%%%%%%%%%%%%%%%%%%%%%%%%%%%%%%%%%%%%%%%%%%%%%%%%%%%
\maketitle
%%%%%%%%%%%%%%%%%%%%%%%%%%%%%%%%%%%%%%%%%%%%%%%%%%%%%%%%%%%%%%%%%
	\section{Introduction}
The first treatment of strong-field ionization of atoms by an alternating electric field was provided in a seminal work by Keldysh \cite{Keldysh65}, who introduced the strong-field approximation (SFA) in a nonrelativistic context. Shortly after, Nikishov and Ritus \cite{Nikishov66,Nikishov67} studied within Klein-Gordon theory the relativistic strong-field ionization of a spinless particle bound by short-range forces in the presence of an electromagnetic plane wave or constant crossed field (CCF). Later on, Reiss \cite{Reiss, Reiss2} generalized Keldysh's consideration to the relativistic regime by calculating, within Dirac theory, the ionization of a 1$s$ electron bound by Coulomb forces in a plane electromagnetic laser wave. In parallel with the SFA approaches, the Perelomov-Popov-Terentev (PPT) theory of strong-field ionization was developed \cite{PPT}. It relies on the imaginary-time method and allows for a relativistic generalization of Keldysh's ionization theory as well \cite{PPT2, Popov98, Popov2006}. Reviews on relativistic strong-field ionization
are given in \cite{ReissGM, Piazza, Krajewska}.\\
\\
Nowadays, it is possible by table-top devices to generate laser pulses with field intensities of the order of $10^{19}-10^{20}$ W/cm$^2$ in the optical or near-infrared frequency domain. The motion of a free electron exposed to such intense fields becomes relativistic within less than an oscillation cycle. Experiments on relativistic strong-field ionization of noble gas atoms have observed the formation of very high charge states with ionization down to the K shell \cite{X0,X,X12,X13}.\\
\\
The rich physics of strong-field ionization is correctly described by the SFA in a qualitative manner. In particular, depending on the value of the Keldysh parameter \cite{Keldysh65}, it properly distinguishes the regimes of multiphoton, above-threshold, and tunneling ionization. The method is also capable of describing  short laser-pulse effects \cite{Kylstra,Krajewska2}. However, to reach predictive power also in quantitative terms, it needs to be adjusted properly. In standard SFA, the ionized electron is described by the corresponding Volkov state, which takes the interaction
with the external electromagnetic (laser) field fully into account but disregards the influence of the atomic core potential on the electron dynamics during tunneling. Accordingly, a correction is needed to incorporate the Coulomb field effects. Corresponding correction factors have been obtained within the framework of the PPT theory \cite{PPT2, Popov98, Popov2006, Popov2005, Brabec, Brabec2}. With respect to the SFA, there are heuristic approaches using Coulomb-Volkov wave functions \cite{Jain}. More recently, it has been shown that the results from the imaginary time method with Coulomb corrections can be derived from the SFA approach using the eikonal-Volkov states for the continuum electron \cite{P1,P2, Krajewska3}. The Coulomb singularity in the phase of the eikonal wave function has been removed, using matching of the continuum wave function to the asymptotics of the undisturbed bound state. More straightforwardly and equivalently, in \cite{Klaiber} it has been shown that the explicit calculation of the Coulomb-corrected SFA amplitude for the atomic $s$-states  avoids singularities because of the cancellation of the Coulomb divergent contribution from the eikonal
phase with a term in the SFA prefactor.  Another method for removing the Coulomb singularity in the SFA amplitude has been put forward in the analytical R-matrix theory, via the imaginary time shifting of the integration in the eikonal phase \cite{ARM1,ARM2}.
\\
\\
It is worth mentioning that, apart from the mainly analytical SFA and PPT approaches, also fully numerical studies of the relativistic dynamics and ionization of highly charged ions in very strong laser fields have been carried out \cite{Y1,Y10, Y100, Y101, Y11,Y12,Y13,Y14}. \\
\\
Due to the approximations involved, SFA predictions turn out to depend on the gauge that is chosen to describe the external electromagnetic fields. From studies of nonrelativistic strong-field ionization it is known that calculations in the length gauge generally yield better agreement with numerical solutions or experimental data than those in velocity gauge \cite{Bauer}. The relativistic version of the length gauge is the Göppert-Mayer gauge \cite{Klaiber, Kylstra, ReissGM}. \\
\\
In the present paper, we provide another treatment of relativistic strong-field ionization of a hydrogen-like atomic system which, to the best of our knowledge, has not been given in the literature yet. As first step, the earlier approaches \cite{Nikishov66,Nikishov67} and \cite{Reiss} are combined to calculate, within Dirac theory and standard SFA, the relativistic ionization of a 1$s$ electron bound by a nuclear Coulomb potential in the presence of a constant crossed field (CCF). In contrast to the work by Nikishov and Ritus \cite{Nikishov66,Nikishov67}, we account for the Coulomb field in the bound state and also for the electron spin. As compared with Reiss \cite{Reiss}, we apply a CCF in Göppert-Mayer gauge, rather than an oscillating wave in radiation gauge. The choice of CCF is motivated by the fact that, for small Keldysh parameters, the ionization resembles a tunneling process through the potential barrier that is formed by the field of the atomic core and the laser field \cite{Klaiber2} which then appears as
quasistatic on the time scale of the ionizaton dynamics. \\
\\
In a second step, we include a suitable Coulomb correction factor which is obtained by an appropriate modification of the procedure developed in \cite{PPT2, Popov98, Popov2006,Popov2005} to the present case. The modification is necessary because the long-range Coulomb potential is already included here in the initial 1$s$ bound state of the electron, whereas a short-range binding potential was assumed in the previous studies \cite{PPT2,Popov98,Popov2006,Popov2005}.\\
\\
Accordingly, our paper is organized as follows. We present our theoretical approach to relativistic strong-field ionization in Sec. II, starting with the standard SFA calculation in Subsec. II.A. In the next subsection II.B we determine an overall Coulomb correction factor that the SFA rate needs to be multiplied with. An analytical approximation of the Coulomb-corrected ionization rate in closed form is obtained in Subsec. II.C. We illustrate our approach by numerical results in Sec. III and compare them with predictions from other theories. Conclusions are given in Sec. IV.
\section{Theoretical framework}
In this section, we outline our approach to relativistic ionization in a CCF in the Göppert-Mayer gauge. In the following, we use atomic units (a.u.) throughout, with the elementary charge unit $e=1$. 
\subsection[A.]{Ionization rate in standard SFA}
The relativistic ionization process in combined laser and Coulomb fields can be described within an S-matrix formalism. The transition amplitude in the SFA can be expressed in the prior form as
\begin{eqnarray}
	S^{\text{SFA}}=-\frac{i}{c}\int \text{d}^4x~ \bar{\psi}^{(-)}(x)\slashed{A}_\text{G}(x)\Phi_{\text{1s}}(x) 
\end{eqnarray}
with the initial state
\begin{eqnarray}
	\Phi_{\text{1s}}(x) = g(r) \chi_s \exp(-i E_\text{1s}t)
\end{eqnarray}
describing the Coulomb-Dirac wave function of the hydrogen like ground state. It consists of a radial part $g(r) = C_\text{1s} (2Zr)^{\sigma-1}\exp(-Zr)$, where $C_\text{1s} = \left( \frac{Z^3}{\pi } \frac{1+\sigma}{\Gamma\left(1+2\sigma\right)}\right)^{\frac{1}{2}}$, the two possible spinors 
\begin{center}
	$\chi_{+\frac{1}{2}}=\begin{pmatrix} 1 \\0 \\ ic \frac{1-\sigma}{Z} \cos\vartheta \\ ic \frac{1-\sigma}{Z} \sin\vartheta \text{e}^{i\varphi} \end{pmatrix}$ and $\chi_{-\frac{1}{2}}=\begin{pmatrix} 0 \\1  \\ ic \frac{1-\sigma}{Z} \sin\vartheta \text{e}^{-i\varphi}\\ - ic \frac{1-\sigma}{Z} \cos\vartheta \end{pmatrix}$
\end{center} 
and the time evolution. Here $E_\text{1s}=\sigma c^2$ with $\sigma = [1-(Z/c)^2]^{1/2}$ indicates the energy of the bound state, $Z$ the nuclear charge number. Correspondingly $I_p=c^2-E_\text{1s}$ denotes the ionization potential. \\
The interaction with the CCF in Eq. (1) is given by the four-potential in the Göppert-Mayer gauge
\begin{eqnarray}
	A_\text{G}^\mu(x)=(-\mathbf{F} \cdot \mathbf{r},-\mathbf{e}_k (\mathbf{F}\cdot \mathbf{r})).
\end{eqnarray}
The coordinate system is oriented with the electric field $\mathbf{F}$ along the $x$-axis and the magnetic field $\mathbf{B}$ along the $y$-axis, with amplitudes $|\mathbf{F}|=|\mathbf{B}|=F$. The CCF can be considered as the infinite-wavelength limit of a plane wave.  In the radiation gauge, it is described by the potential $A^\mu = \tilde{a}^\mu \varphi_k$ with $\tilde{a}^\mu =(0,-a,0,0)$, which is linear in the phase $\varphi_k = k\cdot x$. Note that, when performing calculations in a CCF, it proves useful to introduce a wave four-vector  $k^\mu = \frac{\omega}{c}(1,\mathbf{e}_z)$, along with some frequency $\omega$ as auxiliary quantity. Accordingly, $\mathbf{e}_k = \mathbf{e}_z$ in Eq.~(3). Evidently, all observables must be independent of the parameter $\omega$ at the end. \\
\\
The expression in Eq.~(1) would describe  the exact transition amplitude if the final state accounted for the interaction of the ionized electron with both the CCF and the nuclear Coulomb field. However, such states are not known in analytical form. In  standard SFA, one disregards the influence of the Coulomb field in the continuum state and approximates the latter by a Volkov wave function which is an exact solution of the Dirac equation in a plane-wave-like field. In case of a CCF, it reads  
\begin{eqnarray}
	\psi^{(-)}(x)&=&\sqrt{\frac{c}{p_0}} \left(1 - \frac{\slashed{k}\slashed{A}}{2c(k\cdot p)} \right) u_{p,s} \exp(i S^{(-)}) \notag \\ 
	&  &\times \exp(-i (\tilde{a}\cdot x)\varphi_k/c),  
\end{eqnarray}
with the action
\begin{eqnarray}
	S^{(-)} = -(p\cdot x) + \frac{1}{c(k\cdot p)} \left[\frac{(p\cdot \tilde{a})}{2} \varphi_k^2 - \frac{a^2}{6c} \varphi_k^3 \right], \notag
\end{eqnarray}
a free Dirac spinor $u_{p,s}$, and the asymptotic electron momentum $p^\mu = (p_0, \mathbf{p})$,  where $p_0 = E_p/c$.
The second exponential function in (4) stems from the gauge transformation from the radiation to the Göppert-Mayer gauge. \\
\\
In Eq. (1), following \cite{Nikishov66,Nikishov67}, the term
\begin{eqnarray}
	& &\left(1 - \frac{\slashed{A}\slashed{k}}{2c(k\cdot c)} \right) \notag \\
	&  & \times \exp\left(-i\frac{(p\cdot \tilde{a})}{2c(k\cdot p)} \varphi_k^2 + i\frac{a^2}{6c^2(k\cdot p)} \varphi_k^3  + i\frac{(\tilde{a}\cdot x)}{c} \varphi_k\right)\notag\\
	&= & \int_{-\infty}^{\infty}ds~ \text{e}^{-is\varphi_k}\left(\mathcal{A}(s) + i \frac{\slashed{\tilde{a}}\slashed{k}}{2c(k\cdot p)}\mathcal{A}^\prime(s)\right)
\end{eqnarray}
is expressed by a Fourier integral, where
\begin{eqnarray}
	\mathcal{A}(s)& &= \frac{1}{2\pi}\int_{-\infty}^{\infty}d\varphi_k~ \text{e}^{i\varphi_k s}  \\
	&  &\times \exp\left(-i\frac{(p\cdot \tilde{a})}{2c(k\cdot p)} \varphi_k^2 + i\frac{a^2}{6c^2(k\cdot p)} \varphi_k^3  + i\frac{(\tilde{a}\cdot x)}{c} \varphi_k\right)  \notag
\end{eqnarray}
and the prime denotes the derivative with respect to $s$.
\\
\\
Now the integration over $t$ in Eq.~(1) can be carried out; the resulting $2\pi\delta(E_p - E_\text{1s}-s\omega)$ is afterwards exploited to perform the integration over $s$. \\
\\
For the remaining calculations the vector $\mathbf{q} = \mathbf{p} -s\mathbf{k}- \frac{a}{c}\varphi_k~ \mathbf{e}_x$ and the variables $\alpha = \frac{p\cdot \tilde{a}}{c(k\cdot p)}$, $\beta = \frac{a^2}{8c^2(k\cdot p)}$ and $y = (4\beta)^{2/3}[\frac{s}{4\beta} - (\frac{\alpha}{8\beta})^2]$ with $s=\frac{E_p-E_\text{1s}}{\omega}$ are introduced as well as the substitution $z=(4\beta)^{1/3}(\varphi_k-\frac{\alpha}{8\beta})$ is applied, which leads to 
	\begin{eqnarray}
	S^{\text{SFA}}&= & -\frac{i}{\omega} \sqrt{\frac{c}{p_0}} (4\beta)^{-1/3} \int_{-\infty}^{\infty} dz \int d^3\mathbf{r}~\text{e}^{-i(\mathbf{q}\cdot \mathbf{r})} g(r) \bar{u}_s  \notag \\
	&  &\times \left(1 - \frac{\slashed{\tilde{a}}\slashed{k}}{2c(k\cdot p)} \varphi_k \right) \slashed{A}_\text{G}\chi_s  \\
	&  & \times \exp\left[i\left(yz+\frac{z^3}{3}\right)\right] \exp\left[-i \frac{8\beta}{3}\left(\frac{\alpha}{8\beta}\right)^3+ i \frac{\alpha s}{8\beta}\right]. \notag
\end{eqnarray}

In order to obtain the differential ionization rate, the absolute square of the transition amplitude has to be taken and averaged (summed) over the initial (final) spin polarizations. In carrying out these steps, we use Ref.~\cite{Reiss} as orientation (see also \cite{X2,X21}) but adjust our calculation to the Göppert-Mayer gauge. As a consequence of the latter, the resulting traces over Dirac matrices are simplified decisively, in particular by the fact that the four-products $A_\text{G}^2 (x) = A_\text{G} (x)\cdot A(x) = A_\text{G} (x) \cdot k = A(x)\cdot k = k^2 = A_\text{G} (x)\cdot A_\text{G}(x^\prime)=0$ disappear.  Accordingly $A_\text{G}(x^\prime) = \tilde{a}^\mu \varphi_k(x^\prime) $. The remaining spatial integrations are proportional to 
\begin{eqnarray}
	|S^{\text{SFA}}|^2 &\propto &\int d^3\mathbf{r}\int d^3\mathbf{r}^\prime~ \text{e}^{-i(\mathbf{q}\cdot \mathbf{r})} \text{e}^{i(\mathbf{q}^\prime\cdot \mathbf{r}^\prime)} g(r) g(r^\prime) \notag \\
	&  &\times (c_0+c_3) r^\prime \sin\vartheta^\prime\cos\varphi^\prime r \sin\vartheta\cos\varphi 
\end{eqnarray}
with
\begin{eqnarray}
	c_0&=& 1 +\tau^2[\cos\vartheta\cos\vartheta^\prime+ \sin\vartheta\sin\vartheta^\prime\cos(\varphi-\varphi^\prime)],\notag \\
	c_3&=&i\tau(-\cos\vartheta+\cos\vartheta^\prime)
\end{eqnarray}
and can be calculated straightforwardly in analogy to \cite{Reiss}. Herein $\tau=(1-\sigma)c/Z$. \\
\\
The remaining integrations are of the form
\begin{eqnarray}
	\int dz~\frac{(Z-iQ)^\nu\pm (Z+iQ)^\nu}{(f^\prime(z))^\nu} \exp(-f(z))
\end{eqnarray}
with $f(z) = -i(yz+z^3/3)$, and can be solved by the saddle point method. The physically relevant saddle point is given by $z_0 =i\sqrt{y}$ and the second derivative reads $f^{\prime \prime} (z)=-2iz$. To this end, the denominator is expanded by $f^\prime(z) \approx (z-z_0)f^{\prime\prime}(z_0)$ and the following formula \cite{Gribakin}
\begin{eqnarray}
	\int \frac{\exp(-\lambda f(x))}{(x-x_0)^\nu} dx &\approx& i^\nu \frac{\Gamma(\nu/2)}{2\Gamma(\nu)}\left(\frac{2\pi}{f^{\prime\prime}(x_0)} \right)^{\frac{1}{2}}  \\
	& &\times \left[2\lambda f^{\prime \prime}(x_0) \right]^{\nu/2} \exp[-\lambda f(x_0)] \notag 
\end{eqnarray}
is applied that is valid for large values of $\lambda$. To satisfy the conditions for the applicability of (11), the condition $y\gg1$ must be ensured in Eq. (7). \\
\\
By using the saddle point method, the previously purely real quantity $\mathbf{q}$ becomes complex:
\begin{eqnarray}
	q_x &=& - \frac{az}{c(4\beta)^{1/3}} = - i\frac{\varepsilon}{c}, \notag \\
	q_\perp &=& \sqrt{q_x^2+q_y^2} = i \frac{\zeta}{c}, \notag \\
	q &=& \sqrt{q_x^2+q_y^2+q_z^2} = i \frac{\eta}{c} \notag	
\end{eqnarray}
with the abbreviations
\begin{eqnarray}
	\gamma &=&p_0-p_z, \notag\\
	\eta &=& \sqrt{c^4-E_\text{1s}^2}, \notag \\
	\varepsilon &=& \sqrt{\eta^2+p_y^2c^2 +(c\gamma-E_\text{1s})^2}, \notag \\
	\zeta &=& \sqrt{\eta^2 +(c\gamma-E_\text{1s})^2}.
\end{eqnarray}
The rate is determined by integrating the resulting expression in the common way over the momentum space and by dividing it by the interaction time $T$
\begin{eqnarray}
	R =  \int  \frac{d^3\mathbf{p}}{(2\pi)^3} \frac{|S^\text{SFA}|^2}{T} .
\end{eqnarray}
Noticing that $|S^\text{SFA}|^2$ does not depend on $p_x$, the corresponding integral in Eq.~(13) yields
\begin{eqnarray}
	\int dp_x = FT
\end{eqnarray}
(see also \cite{Nikishov66,Nikishov1}). The remaining $p_y$- and $p_z$-integrations are transformed into integrations over $p_y$ and $\gamma$, according to
\begin{eqnarray}
	R &=& \frac{F^2  C_\text{1s}^2}{4} (2Z)^{2\sigma-2} \int_{-\infty}^{\infty} dp_y \int_{0}^{\infty}d\gamma  \frac{\gamma}{\varepsilon} \text{e}^{-\frac{4}{3} y^\frac{3}{2}}\bigg[ |\tilde{S}_1(p_y,\gamma)|^2\notag \\
	&  &+ \tau^2 \left( |\tilde{S}_2(p_y,\gamma)|^2+ |\tilde{S}_3(p_y,\gamma)|^2 +|\tilde{S}_4(p_y,\gamma)|^2 \right) \notag \\
	&  &+2 \tau\text{Re}\left[\tilde{S}_1(p_y,\gamma)\right]\text{Im}\left[\tilde{S}_2(p_y,\gamma)\right]\bigg] 
\end{eqnarray}
with 
\begin{eqnarray}
	\tilde{S}_1(p_y,\gamma) &=&   \varepsilon \left[ \frac{ c^2}{\eta^3} \Gamma(\frac{\sigma +1}{2}) D^{\sigma+1} - \frac{c}{\eta ^2} \Gamma(\frac{\sigma +2}{2}) D^{\sigma+2}\right], \notag \\
	\tilde{S}_2(p_y,\gamma) &=& i  \varepsilon ~q_z \bigg[ 3 \frac{ c^4}{\eta^5} \Gamma(\frac{\sigma}{2}) D^\sigma - 3 \frac{ c^3}{\eta^4} \Gamma(\frac{\sigma+1}{2})D^{\sigma+1} \notag \\
	& & +\frac{ c^2}{\eta^3} \Gamma(\frac{\sigma+2}{2})D^{\sigma+2} \bigg], \notag \\
	\tilde{S}_3(p_y,\gamma) &=& - \frac{\varepsilon}{\zeta} \bigg[ \left(3\frac{\zeta^2c^3}{\eta^5}- \frac{c^3}{\eta^3}\right) \Gamma(\frac{\sigma}{2})D^\sigma \notag \\
	&  &+ \left( \frac{c^2}{\eta^2}-3\frac{\zeta^2c^2}{\eta^4}\right)\Gamma(\frac{\sigma+1}{2})D^{\sigma+1} \notag \\
	&  &+ \frac{\zeta^2 c}{\eta^3}\Gamma(\frac{\sigma+2}{2})D^{\sigma+2}\bigg],\notag \\
	\tilde{S}_4(p_y,\gamma) &=& -i\frac{p_y c}{\zeta}\left[ \frac{c^3}{\eta^3} \Gamma(\frac{\sigma}{2})D^\sigma -\frac{c^2}{\eta^2} \Gamma(\frac{\sigma+1}{2})A^{\sigma+1}\right] \notag
\end{eqnarray}
and $D = Zc \left(\frac{2}{F\gamma\varepsilon} \right)^\frac{1}{2}$. \\
\\
The remaining integrations can be carried out by numerical means without difficulties.

\subsection[B.]{Coulomb correction}
Up to this point, the influence of the Coulomb field on the continuum state has not been considered. However, especially during the tunneling process, the electron continues to experience the influence of this field. Therefore, what is needed is a way to account for the effect of the nucleus on the ionized electron. For an electron bound by a short-range potential, this problem has already been widely treated \cite{PPT2, Popov98, Popov2006,Popov2005,Brabec,Brabec2}. Accordingly, to include the Coulomb interaction between the atomic core and the escaping electron in this case, its influence is taken into account under the complete tunneling barrier. This is achieved by dividing the barrier length into rather close distances, where the Coulomb effects can be incorporated via a typical Coulomb logarithm in the wave function, and rather large distances where the Coulomb potential can be treated as a perturbation, with both regions being linked through a matching procedure \cite{PPT2, Popov98, Popov2006,Popov2005,Brabec,Brabec2}.\\
\\ 
In our case, however, we do not assume short-range binding forces but, instead, fully account for the long-range nuclear Coulomb field in in the bound 1$s$ state. A part of the total Coulomb effects is therefore already contained in Eq.(15). As a consequence, we have to appropriately 'truncate' the Coulomb corrections established in \cite{PPT2, Popov98, Popov2006,Popov2005,Brabec,Brabec2} in order to avoid their overestimation. To this end, we restrict the influence on the escaping electron to the region in which the Coulomb field, as compared with the impact of the CCF, is a perturbation. It will turn out that this procedure provides a remarkably good approximation to the Coulomb effects. \\
\\
We determine the Coulomb correction as \cite{PPT2, Popov98, Popov2006,Popov2005,Brabec,Brabec2}
\begin{eqnarray}
	Q &= &\exp\bigg(2iZ \int_{t_1}^{0}\frac{1}{r(t)}dt \bigg),
\end{eqnarray}
where $t=t_1$ denotes the time at which the Coulomb  potential starts to be taken into account and $t=0$ denotes the time of the tunnel exit. \\
The trajectory $r(t)$ is obtained by a solution of the classical, relativistic equations of motion in a CCF, which is adapted to the initial conditions $\vec{r}(t_0)=0, \text{Im}(\vec{r}(0))=0, \text{Im}(\dot{\vec{r}}(0))=0$ and $\frac{1}{\sqrt{1-\dot{\vec{r}}^2(t_0)/c^2}}=\epsilon_0$, with $\epsilon_0=E_\text{1s}/c^2$. Here $t=t_0$ is the time of tunnel entry. This results in the parameterized coordinates \cite{PPT2,Brabec,Brabec2}
\begin{eqnarray}
	x(u)&=&\frac{c^2}{2F\lambda}(u^2-u_0^2), y(u)=0, \notag \\
	z(u)&=&i\frac{c^2}{6F\lambda}(u^2-u_0^2)u, \notag \\
	ct(u)&=&i\frac{c^2}{6F\lambda}u^3-i\frac{c^2}{2F\lambda}(1+\lambda^2)u.
\end{eqnarray}
with $u_0^2=3(\lambda^2-1)$ and $\lambda = -\frac{\epsilon_0}{2}+\frac{1}{2}\sqrt{\epsilon_0^2+8}$.\\
\\
An intuitive approach to determine $t_1$ is motivated by the idea that for $r_1 = r(t_1)$ the Coulomb field strength equals the CCF amplitude and, afterwards, falls below it. This leads to 
%We determine the start time $t_1$ in Eq.~(17) by the physically motivated requirement that, at $r_1 := r(t_1)$, the Coulomb field strength equals the CCF amplitude and, afterwards, falls below it. This yields
\begin{eqnarray}
	 r_1 &=& \sqrt{\frac{Z}{F}}. 
\end{eqnarray}
We note that $r_1$, while being substantially larger than the size $\sim 1/Z$ of the bound state, is much smaller than the extension $\sim Z^2/F$ of the tunneling barrier. From this, $t_1$ and $u_1$ can be determined:
\begin{eqnarray}
	\frac{c^2}{2F\lambda}(u_0^2-u^2)\sqrt{1-\frac{u^2}{9}} &=& r_1 .
\end{eqnarray}
Taking the tunnel distance as a pure $x$ component, where $|x(u)| \gg |z(u)|$ holds, the equation is simplified to
\begin{eqnarray}
  u_1 &=& \sqrt{u_0^2 -\frac{2F^{\frac{1}{2}} Z^{\frac{1}{2}}\lambda }{  c^2}}. 
\end{eqnarray}
Interestingly, the intuitive value of $r_1$ (respectively the associated coordinate $x_1$) given above can be mathematically supported and further improved by noting that it conincides approximately with the saddle point of the integrand in the S-matrix element of Eq.(1). Let us briefly sketch the corresponding derivation\cite{SPM}. The space-time dependence of the integrand in cylindrical coordinates is given by 
\begin{eqnarray}
	m &=& \rho^2 \cos(\phi) \text{e}^{-i S[\rho, \phi, z, t]} (2 Z 
	\sqrt{\rho^2 + z^2})^{\sigma - 1} \notag \\
	&  & \times \text{e}^{-Z\sqrt{\rho^2 + z^2}}
	\text{e}^{-i \sigma c^2 t} 
\end{eqnarray}
where $S[\rho, \phi, z, t]$ denotes the classical action, which satisfies the Hamilton-Jacobi equation 
\begin{eqnarray}
	\left(\nabla S + \frac{1}{c} \mathbf{A}_\text{G} \right)^2-\frac{1}{c^2}\left(\frac{\partial S}{\partial t} - A_\text{G}^0-\frac{Z}{r}\right)^2 +c^2=0.~~
\end{eqnarray}
Note that the angular dependence of the spinors $\chi_{\pm 1/2}$ has been omitted in Eq.~(21), as it is contained solely in the lower components which are suppressed by the small parameter $\tau$. We can now substitute the saddle point conditions into this equation and obtain an equation for $x$, which can be solved approximately by exploiting the smallness of $F/F_a$  (with the atomic field strength $F_a = Z^3$) and $I_p/c^2$. With next-to-leading order accuracy, one finds
\begin{eqnarray}
	\tilde{x}_1= -\sqrt{\frac{Z}{F}} \left(1 +  \frac{Z^2}{9c^2}\right) + \frac{1}{Z} 
\end{eqnarray}
and
\begin{eqnarray}
	\tilde{u}_1 &=& \sqrt{u_0^2 + \frac{2F\lambda}{  c^2} \tilde{x}_1}. 
\end{eqnarray}
We point out that, here we have taken the influence of the $z$-component into account, which modifies the second term in the brackets in Eq.(23). The Coulomb correction is, accordingly, determined to
%In the non-relativistic parameter regime, we can show that the coordinate saddle point is indeed at sqrt(Z/E0) and that the first correction is $-1/Z$. To do this, we calculate $\int dx dt \exp(-iS(x,t))Hi(x,t)phi(x,t)$ without knowledge of the exact classical action in Coulomb and laser field using saddle point integration. $S(x,t)$ satisfies the Hamilton-Jacobi equation and one can now substitute the saddle point conditions into this equation and obtain an equation for $x$. For the relativistic generalization, we start the same, but use the form of the integrals averaged over the spin states at leading order.\\
\begin{eqnarray}
	Q & =& \left(\frac{\sin(\varphi_0 + \varphi_1)}{\sin(\varphi_0 - \varphi_1)}\right)^{2\beta} \exp\left[\frac{6Z\varphi_1}{c}\right] 
\end{eqnarray}
with $\beta =Z\epsilon_0(1-\epsilon_0^2)^{-1/2}/c$, $\varphi_0=\arcsin(u_0/3)$ and $\varphi_1 = \arcsin(\tilde{u}_1/3)$.
\begin{figure}[htp]
	\centering
	\includegraphics[width=0.95\linewidth]{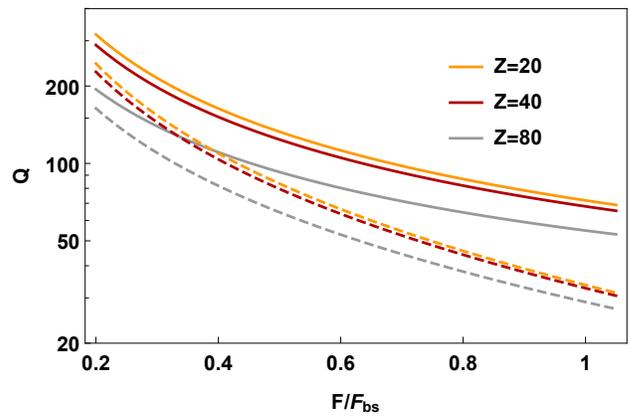}
	\caption{Coulomb correction factor from Eq.~(25) as function of the normalized field strength $F/F_{bs}$, where $F_{bs} = c^4 (1-\sigma)^2 / (4 Z)$ denotes the barrier suppression field strength. For comparison the Coulomb correction factor for the intuitive approach with $\varphi_1 = \arcsin(u_1/3)$ [see Eq. (20)] is shown as dashed curves.}
\end{figure}
\\
For illustration, Fig. 1 shows the Coulomb correction factor from Eq. (25) by solid lines and the intuitive correction factor based on Eq. (20) by dashed lines. The more accurate description yields larger corrections, especially at high field strength. Compared to the well-known result of \cite{PPT2}, however, the present Coulomb correction factors turn out to be much smaller (by about two orders of magnitude). As mentioned before, this is because in \cite{PPT2} the correction is evaluated along the complete tunnel length.\\

\subsection[C.]{Analytical simplifications}
The fact that the prefactor of the exponential dependency $\text{e}^{-\frac{4}{3} y^\frac{3}{2}}$ in Eq. (15) changes only very slowly as a function of $\gamma$ and $p_y$ can now be used to solve the remaining integrations. The $p_y$-dependence is contained in the variables $\varepsilon$ and $\zeta$, as well as in the exponential dependence by the variable $y$. In order to be able to carry out the integration over $p_y$ the prefactor is evaluated at $p_y=0$ and the function in the exponential is expanded around $p_y=0$ up to the second order. The $p_y$ dependency in Eq.~(15) can then be evaluated analytically as a Gaussian integral to very good approximation. \\
\begin{figure}[htp]
	\centering
	\includegraphics[width=0.95\linewidth]{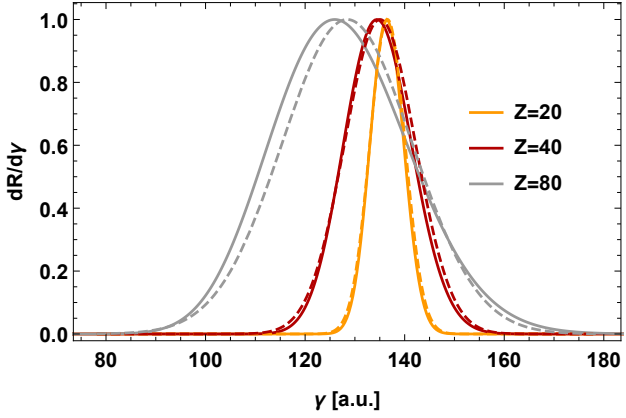}
	\caption{Differential rate $\frac{dR}{d\gamma}$ as a function of $\gamma$ for different nuclear charge numbers $Z$ at $F=F_{bs}$ (solid lines), when the $p_y$-dependence in Eq. (15) has been integrated numerically. The dashed lines show the approximation when the exponential prefactor in Eq.~(15) is evaluated at $p_y=0$ and $\gamma=\gamma_0$, the $p_y$-dependence is integrated out analytically and the remaining exponent is expanded up to second order. The curves have been normalized to a height of 1 to facilitate their comparison.}
\end{figure}
\\
The $\gamma$-dependence is far more complex. Nevertheless, it proves to be practicable to integrate only over the exponential dependence. Therefore the value $\gamma_0$ is determined at which the exponential function reaches its maximum:
\begin{eqnarray}
	\gamma_0 &=& \frac{E_\text{1s}}{4c}+ \frac{1}{4c}\sqrt{E_\text{1s}^2+8c^4}.
\end{eqnarray}
The exponent is expanded up to the second order around $\gamma_0$ and the prefactor is evaluated at this  point. That this approach reflects the differential rate from Eq.~(15) very well, can be seen in Fig.~2. Here, $dR/d\gamma$ from Eq.~(15) is shown as a solid line for different nuclear charge numbers $Z$. In comparison, the approach described above by evaluating the prefactor at $\gamma_0$ and expanding the exponent is shown as a dashed line. The maximum of the exact differential rate lies only slightly below $\gamma_0$ and the widths of both curves (for given $Z$) is almost the same. By formally extending the lower integration boundary to $-\infty$, the $\gamma$ integration over the approximated expression for $dR/d\gamma$ can be performed as a Gaussian integral as well. This results in
\begin{eqnarray}
	R &=& \frac{1}{4} \pi F^2  C_\text{1s}^2 (2Z)^{2\sigma-2} \frac{\gamma_0}{\varepsilon_0} \sqrt{\frac{ 2 F \gamma_0}{\varepsilon_0 h^{\prime \prime}(\gamma_0)}} \text{e}^{-h(\gamma_0)}  \\
	& & \times \bigg[ |\tilde{S}_1(p_y=0,\gamma=\gamma_0)|^2\notag \\
	&  &+ \tau^2 \left( |\tilde{S}_2(p_y=0,\gamma=\gamma_0)|^2+ |\tilde{S}_3(p_y=0,\gamma=\gamma_0)|^2  \right) \notag \\
	&  &+2 \tau\text{Re}\left[\tilde{S}_1(p_y=0,\gamma=\gamma_0)\right]\text{Im}\left[\tilde{S}_2(p_y=0,\gamma=\gamma_0)\right]\bigg] \notag 
\end{eqnarray}
with 
\begin{eqnarray}
	\varepsilon_0 &=& \sqrt{\eta^2 +(c\gamma_0-E_\text{1s})^2}, \notag \\
	h(\gamma) & = & \frac{2\varepsilon^3}{3 c^2 F \gamma}. \notag 
\end{eqnarray}
To further simplify the expression, $\gamma_0$ can be expanded for small values of $I_p/c^2\ll 1$ to give
\begin{eqnarray}
	\gamma_0 & \approx& c - \frac{I_p}{3c} .
\end{eqnarray}
Accordingly, the main contribution of $p_z$ is around $\frac{I_p}{3c}$. This behavior is in agreement with the result of \cite{Klaiber} where the same shift of the momentum distribution along the $z$ resp. $k$ direction has been found. \\
\\
By noticing that $\tilde{S}_1 \gg \tau \tilde{S}_i$ for $ i \in \{2,3,4\} $ in the relevant range of parameters and by expanding all quantities up to the first order in $I_p/c^2\ll 1$, the rate can be approximately written as
\begin{eqnarray}
	R &\approx&  \frac{F^{2-\sigma}F_a^{1+\frac{4\sigma}{3}}}{(2I_p)^{\frac{7}{2}+\frac{\sigma}{2}}}\frac{1+\sigma}{\Gamma(1+2\sigma)} 2^{3\sigma-3} \frac{1-\frac{7}{72}\frac{I_p}{c^2}}{\sqrt{1 +\frac{5}{12} \frac{I_p}{c^2}}} \notag \\
	&  & \times \frac{1}{(1-\frac{1}{2}\frac{I_p}{c^2})^3} \Gamma(\frac{\sigma+1}{2})^2  \left[ 1+\frac{17}{36}\frac{I_p}{c^2}  \right]^{\sigma} \notag \\
	&  & \times \left[1 - \frac{\Gamma(\frac{\sigma +2}{2})}{\Gamma(\frac{\sigma +1}{2})} Z\bigg(2\frac{\sqrt{2I_p}}{F} (1-\frac{1}{36} \frac{I_p}{c^2}) \bigg)^{\frac{1}{2}} \right]^2 \notag \\
	&  & \times \exp\bigg[- \frac{2}{3}\frac{(2I_p)^{\frac{3}{2}}}{F} \left(1-\frac{1}{12}\frac{I_p}{c^2}\right)\bigg] . 
\end{eqnarray}
with the atomic field strength $F_a=Z^3$. While the exponential characteristic of the ionization process ~$ \exp[-\frac{2}{3}\frac{(2I_p)^{\frac{3}{2}}}{F}]$ is generally found in tunneling-like rate expressions \cite{Keldysh65,Nikishov66,Nikishov67,PPT2, Popov98, Popov2006,Popov2005,Brabec, Brabec2}, it is particularly noteworthy that the additional term $I_p/12c^2$ in the last line of Eq. (29)  coincides as well with that of Ref.\cite{Klaiber}. The various relativistic ionization theories are known to somewhat differ in the nuclear charge and field strength dependencies of the pre-exponential factors they predict.\\
\\
To account for the Coulomb correction, the standard SFA-rate from Eq.~(15) resp. (27) has to be multiplied with the factor $Q$:
\begin{eqnarray}
	R_Q =R\cdot Q.
\end{eqnarray}
In the next section the rate $R_Q$ is compared with previously existing calculations and it is placed in context.
\section{Results and Discussion}
\begin{figure}[htp]
	\centering
	\includegraphics[width=0.95\linewidth]{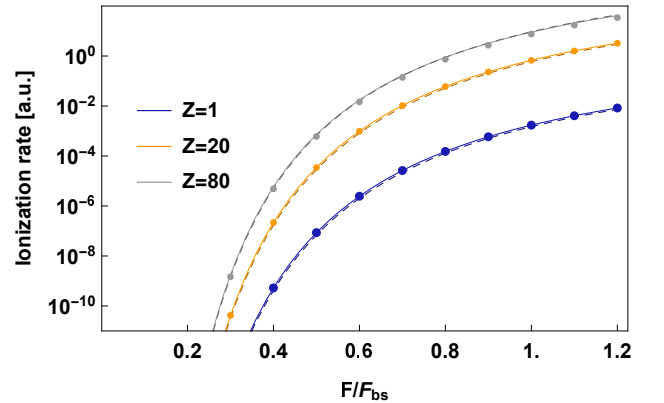}
	\caption{Total ionization rate as function of the normalized field strength $F/F_{bs}$. The solid curves show Eq.(27) corrected by the Coulomb factor (25), the corresponding analytical approximation (29) including the same Coulomb factor is depicted by dots and the Coulomb corrected rate of \cite{Brabec, Brabec2} is shown as dashed curves.}
\end{figure}
In this section, we illustrate our theoretical approach by showing the resulting ionization rates in a wide range of field strengths and ionic systems. Moreover, we compare our findings with already existing calculations of relativistic strong-field ionization.\\
\\
The dependence on the field strength is mainly determined by the exponential dependence of the rate. Different ionization theories can, however, differ in the preexponential factors they predict. \\
Figure 3 shows the total ionization rate (27) corrected by the Coulomb factor (25) as a function of the normalized field strength (solid curves). The steep exponential dependence is clearly seen. For comparison the PPT result of \cite{Brabec,Brabec2} is presented as dashed curves. On the displayed logarithmic scale, our results lie almost completely on the PPT curves and differ only slightly.\\
The dotted curves give the Coulomb corrected Eq.~ (29). The approximation coincides very well with the numerically calculated solid curves and deviates only slightly from this result for large nuclear charge numbers $Z$ and large field strengths.\\
\\
\begin{figure}[h]
	\centering
	\subfloat[][]{\includegraphics[width=0.9\linewidth]{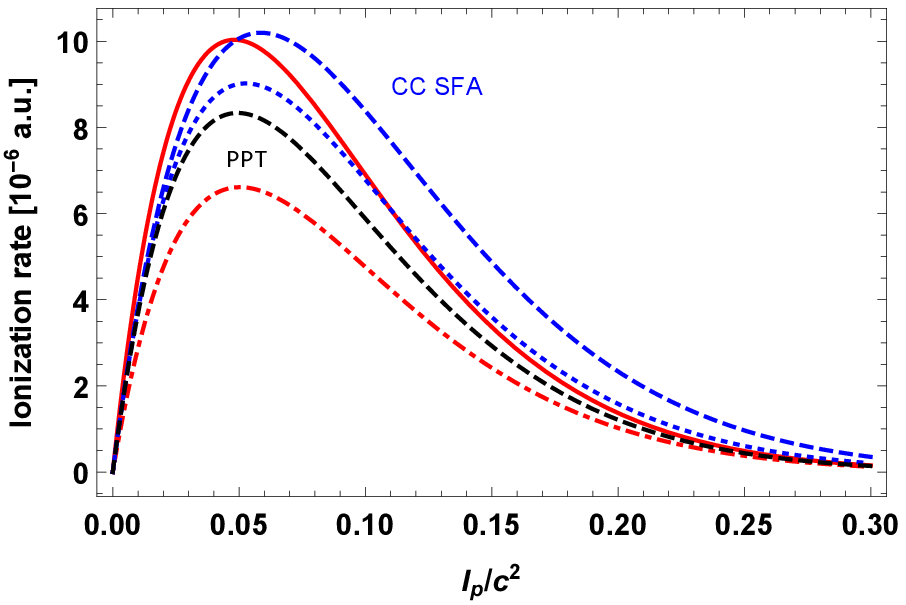}}
	\qquad
	\subfloat[][]{\includegraphics[width=0.9\linewidth]{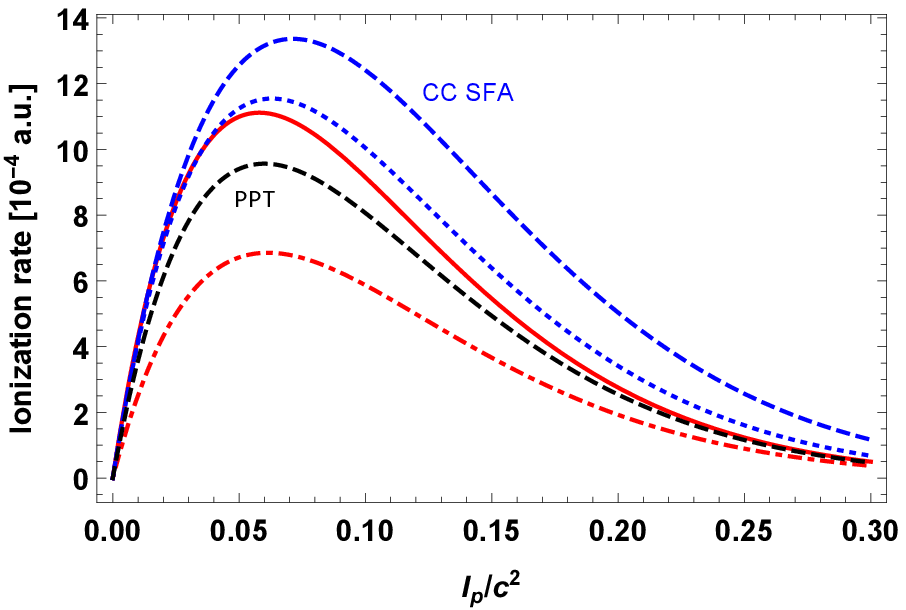}}
	\caption{Total ionization rate as function of the normalized ionization potential $I_p/c^2$ for (a) $F/F_a=1/32$ and (b) $F/F_a=1/25$. The red solid lines show the -- over half a cycle of an oscillating laser wave averaged -- results from Eq.~(27), including the Coulomb correction factor (25). The red dotdashed lines show the result for the intuitive Coulomb correction factor with $\varphi_1 = \arcsin(u_1/3)$ [see Eq. (20)]. The blue dashed (dotted) curves represent Eq. (78) (Eq. (97)) from \cite{Klaiber} and the black dashed lines shows the cycle-averaged result of \cite{Brabec,Brabec2}.}
\end{figure}
%\begin{figure}[htp]
%	\centering
%	\includegraphics[width=0.95\linewidth]{VergleichKlaiber.eps}
%	\caption{Total ionization rate as function of the normalized ionization potential $I_p/c^2$ for $F/F_a=1/32$. The red solid and red dashed lines show the -- over half a cycle of an oscillating laser wave averaged -- results from Eq.~(15) and (25), respectively, including the Coulomb correction factor (21). The brown curve represent equation (97) from \cite{Klaiber} and the black line shows the cycle-averaged result of \cite{Brabec,Brabec2}.}
%\end{figure}
The dependence of the ionization rate on the nuclear charge, which is encoded by the normalized ionization potential, is shown in Fig. 4 on a linear scale. Our rate in a CCF has been averaged over half a cycle of an oscillating laser wave with electric field of the form $F \sin(\varphi_k)$. It is compared with the PPT results from Ref.~\cite{Brabec,Brabec2}  (also averaged over half a laser period) and the SFA results from Ref.~\cite{Klaiber}, where ionization by a linearly polarized laser wave was considered. We note in this regard, that Eq. (78) and (97) in Ref. \cite{Klaiber} represent two variants of the Coulomb-corrected SFA that are obtained from different partitions of the underlying Hamiltonian. While Eq. (78) relies on the standard partition, in the 'dressed' Coulomb-corrected SFA of Eq. (97) the laser field makes some contribution to the bound state evolution. The latter brings the rates closer to the PPT prediction. When we apply our intuitive Coulomb correction factor based on Eq.(20), the ionization rate comes out a bit too low, as the red dashed curves show. This correction factor, thus, slightly underestimates the Coulomb effects. The shape of the curves is, however, very similar to the PPT result of \cite{Brabec,Brabec2} and Eq.(97) from \cite{Klaiber}. Applying the more accurate Coulomb correction (25) improves the ionization rates significantly (red solid curves). Now they resemble more closely the predictions from \cite{Klaiber} that include the Coulomb effects in a coherent manner by using eikonal-Volkov states. The remaining differences as compared with this advanced SFA theory indicate that further correction terms (especially those stemming from the innermost region of the tunnel) would be needed in our approach to enhance the agreement.

\section{Conclusion}
An alternative treatment of relativistic
strong-field ionization of hydrogen-like atomic systems has been presented. Our approach combines various calculational methods that have been developed in earlier investigations of the problem \cite{Nikishov66,Nikishov67,Reiss,PPT2}. First, we have calculated the total ionization rate of a 1$s$ electron bound by a nuclear Coulomb potential in the presence of a constant crossed field in Göppert-Mayer gauge within Dirac theory and standard SFA. To take the influence of the Coulomb field on the electron continuum state during tunneling into consideration, a well-known method that relies on the assumption of a short-range binding potential \cite{PPT2}, has been adjusted appropriately. Two versions of the resulting modified Coulomb correction factor have been obtained, an intuitive one and a more accurate one, that are effectively active in the region where the nuclear Coulomb field has fallen below the CCF. They may be interpreted as arising from the necessity to incorporate Coulomb effects into the Volkov state. We emphasize that both versions of our Coulomb correction factor represent approximations as they disregard the innermost subbarrier region.\\
\\
We have derived a representation of the Coulomb-corrected total ionization rate as double integral, which could be very well approximated by an analytical formula in closed form. Comparing our corresponding results with predictions from previous studies based on PPT \cite{Brabec, Brabec2} or Coulomb-corrected SFA \cite{Klaiber} theories, we found good agreement in a wide range of nuclear charges (resp. ionization potentials) for field strengths below the barrier suppression field. Our study may offer additional insights into relativistic strong-field ionization, in general, and the role of Coulomb corrections, in particular. It may, moreover, help to establish further connections between corresponding PPT and SFA theories.
\\
\\
As outlook we note that our approach could be appropriately applied to other strong-field problems as well to obtain, for example, a Coulomb-corrected rate for bound free pair production in an intense laser field \cite{X2,X21}.
\\
\begin{acknowledgements}
	This work has been funded by the Deutsche Forschungsgemeinschaft (DFG) under Grant No. 392856280 within the Research Unit FOR 2783/1. Useful discussions with K.~Z. Hatsagortsyan are gratefully acknowledged.
\end{acknowledgements}

\end{document}